\documentclass[12pt,preprint]{aastex}
\def\t0{\theta_{\circ}}

\def\be{\begin{equation}}
\def\en{\end{equation}}

\newcommand{\msun}{$M_{\odot}$}
\newcommand{\mdot}{$\dot{M}$}

\begin{document}


\title{Evidence for a T Tauri Phase in Young Brown Dwarfs}

\author{Ray Jayawardhana}
\affil{Department of Astronomy, University of Michigan, 830 Dennison, Ann Arbor, MI 48109}

\author{Subhanjoy Mohanty}
\affil{Harvard-Smithsonian Center for Astrophysics, 60 Garden St., Cambridge, MA 02138}

\and

\author{Gibor Basri}
\affil{Department of Astronomy, University of California at Berkeley, Berkeley, CA 94720}

\begin{abstract}
As part of a multi-faceted program to investigate the origin and early 
evolution of sub-stellar objects, we present high-resolution Keck optical 
spectra of 14 very low mass sources in the IC 348 young cluster and the 
Taurus star-forming cloud. All of our targets, which span a range of 
spectral types from M5 to M8, exhibit moderate to very strong H$\alpha$ 
emission. In half of the IC 348 objects, the H$\alpha$ profiles are broad 
and asymmetric, indicative of on-going accretion. Of these, IC348-355 (M8) 
is the lowest mass object to date to show accretion-like H$\alpha$. 
Three of our $\sim$ M6 IC 348 targets with broad H$\alpha$ also harbor 
broad OI (8446\AA) and CaII (8662\AA) emission, and one 
shows broad HeI (6678\AA) emission; these features are usually seen in 
strongly accreting classical T Tauri stars.  We find that in very low mass 
accretors, the H$\alpha$ profile may be somewhat narrower than that in higher 
mass stars. We propose that low accretion rates combined with small infall 
velocities at very low masses can conspire to produce this effect. In the 
non-accretors in our sample, H$\alpha$ emission is commensurate with, or 
higher than, saturated levels in field M dwarfs of similar spectral 
type.  Our results constitute the most compelling evidence to date that 
young brown dwarfs undergo a T Tauri-like accretion phase similar to that 
in stars. This is consistent with a common origin for most low-mass stars, 
brown dwarfs and isolated planetary mass objects.
\end{abstract}

\keywords{stars: low mass, brown dwarfs -- stars: pre-main-sequence -- 
circumstellar matter -- planetary systems -- stars: formation -- 
techniques: spectroscopic}

\section{Introduction}
Brown dwarfs, which straddle the mass range between stars and planets, 
appear to be common both in the field and in star-forming regions. Their
ubiquity makes the question of their origin an important one, both for 
our understanding of brown dwarfs themselves as well as for theories on 
the formation of stars and planets. 

In the standard framework, a low-mass star forms out of a collapsing 
cloud fragment, and goes through a ``T Tauri phase'', during which it 
accretes material from a surrounding disk, before arriving on the main 
sequence. There is ample observational evidence now to support many key
aspects of this picture for young solar-mass stars. One of the defining 
characteristics of the T Tauri phase is a rich array of emission lines 
in the spectra, of which H$\alpha$ is usually the most prominent (Joy 
1945). The broad, asymmetric H$\alpha$ line profiles, seen
in ``classical'' T Tauri stars, are now thought to arise in the infall 
region, as are many of the metallic emission lines (Muzerolle, Hartmann 
\& Calvet 1998 and references therein). 

Whether the same scenario holds for objects at and below the sub-stellar 
limit is an open question. It has been suggested, most recently by Padoan 
\& Nordlund (2003), that brown dwarfs form in the same way as more massive 
stars, via `turbulent fragmentation'. Reipurth \& Clarke (2001) proposed 
an alternate scenario, which has been further explored through numerical 
simulations by Bate et al. (2002; 2003): in their model, brown dwarfs (and 
presumably isolated planetary mass objects or ``planemos'') are stellar 
embryos, ejected from newborn multiple systems before they can accrete 
sufficient mass to eventually fuse hydrogen. In this scenario, a 
stellar embryo competes with its siblings in order to accrete infalling 
matter, and the one that grows slowest is most likely to be ejected through
dynamical interactions. A key prediction of Reipurth \& Clarke (2001) is 
that ``sub-stellar equivalents to the classical T Tauri stars should be 
rather short-lived'' because the ejection process limits the amount of gas 
brought along in a disk. 

Studies of {\it young} sub-stellar objects could provide valuable clues 
to distinguish between these formation mechanisms. Therefore, we have 
undertaken a multi-faceted study of very low mass (VLM) objects in 
star-forming regions and their immediate circumstellar environment. One 
of our key goals is to investigate whether some or all young brown dwarfs 
undergo a T Tauri-like phase, and if so how long that phase lasts. 

This paper reports on a search for accretion signatures in high-resolution 
spectra of 14 objects near and below the sub-stellar boundary in IC 348 
($\sim$ 320 pc) and Taurus ($\sim$ 150 pc). Our targets consist 
of M5--M8 sources in IC 348 from Luhman (1999) and the recently identified 
M7--M8 brown dwarfs in Taurus from Mart\'\i n et al. (2001)\footnote{The 
spectral classification of the Taurus targets by Mart\'\i n et al. (2001) 
has been modified by Brice\~{n}o et al. (2002).  We adopt the latter authors' 
types, which are consistent with Luhman's (1999) typing scheme for the 
IC 348 targets.}. Combined with the spectra of 15 additional sources in 
the Upper Scorpius and $\rho$ Ophiuchus star-forming regions (Jayawardhana, 
Mohanty \& Basri 2002), we now have a substantial sample to characterize 
disk accretion in young brown dwarfs. 

In a subsequent paper, we will use these optical spectra to derive 
temperatures, gravities and masses for the current 
sample, as we did for Upper Sco and $\rho$ Oph (Mohanty et al. 2003a, 2003b). 
For now, we assume an age of $\sim$ 1--2 Myr for IC 348 (Luhman et al. 2000) 
and $\sim$ 3 Myr for Taurus (White \& Ghez 2001). If we further adopt the 
Luhman (1999) spectral type-temperature scale for low-mass pre-main sequence 
objects, then theoretical evolutionary models (Baraffe et al. 1998) imply 
masses of $\sim$ 0.1 -- 0.02 \msun~ for our M5-M8 objects.  Thus, our sample 
spans a mass range from the stellar/sub-stellar boundary well into the brown 
dwarf domain (and possibly even into the planetary mass regime; Mohanty 
et al. 2003b).

\section{Observations and Analysis}
We obtained optical spectra of the target sample using the High Resolution 
Echelle Spectrometer (HIRES; Vogt et al. 1994) on the Keck I telescope on
2001 October 27 UT (CFHT-BD-Tau-2, 3 and 4) and on 2002 October 31 UT 
(the rest, plus a second spectrum of CFHT-3). With the 1.15'' slit, the two-pixel-binned spectral resolution 
is R $\approx$ 33,000. The instrument yielded 15 spectral orders in the 
6390 -- 8700 \AA~ wavelength range, with gaps between the orders,
providing a variety of features related to youth and accretion activity. 
For comparison to our targets and to derive {\it v~sin~i}, we used M dwarf
and M giant spectroscopic standards observed with the same HIRES set up. 
The data were reduced in a standard manner using IDL routines, as described 
in Basri et al. (2000). 

We derived rotational velocities ($v$~sin~$i$) of the targets by 
cross-correlating with a `spun-up' template of a slowly rotating standard.
Multiple spectral orders ($\sim$ 6), selected on the basis of an absence of
strong telluric features, strong gravity-sensitive features, and stellar 
emission lines, were used in the cross-correlation analysis. Following 
White \& Basri (2003), we used a combination of giant and dwarf spectra for 
the template (see detailed discussion in Mohanty \& Basri 2003 and Mohanty et 
al. 2003a). 

\section{Results and Discussion}
Table 1 lists the derived $v$~sin~$i$, H$\alpha$ equivalent widths, and 
H$\alpha$ full widths at 10\% of the peak flux for our sample.  Figure 1 
shows H$\alpha$ line profiles in our sample. Of the ten IC 348 objects, five 
show broad, asymmetric H$\alpha$ lines;
three of those five also exhibit strong emission in OI (8446\AA) and CaII 
(8662\AA) (Fig. 2). Of the four Taurus targets, CFHT-BD-Tau-3 and 
CFHT-BD-Tau-4 show very strong but relatively narrow H$\alpha$.  We have 
clear detections of Li (6708 \AA) absorption in IC348-256, 286 and 
353, and marginal detections in CFHT-BD-Tau-2 and 3.  The other spectra 
are too noisy in this region for unambiguous detection (not surprising for 
these very faint, red objects). 

\subsection{Disk Accretion}
The shape and width of the H$\alpha$ emission profile is commonly used to 
discriminate between accretors and non-accretors among T Tauri stars (TTS). 
Stars exhibiting broad, asymmetric H$\alpha$ lines with equivalent width 
larger than 10 \AA~ are generally categorized as classical TTS (CTTS), 
although this threshold value varies with spectral type (e.g., Mart\'in 1998).
Recently White \& Basri (2003; hereafter WB) have suggested that a full-width 
$>$ 270 kms$^{-1}$, at 10\% of the peak emission, is a better empirical 
indicator of accretion, independent of spectral type.  

By the latter criterion, two of our targets are accretors: IC348-165 (M5.25) 
and IC348-205 (M6), with 10\% full-width $>$ 300 kms$^{-1}$.  Moreover, both 
objects show broad (FWHM $>$ 100 kms$^{-1}$) emission in OI (8446\AA) and CaII
(8662\AA); IC348-165 also exhibits broad emission in HeI (6678\AA).  In CTTS, 
the presence of a broad component in these metallic emission lines is 
associated with moderate to high accretion rates (\mdot $\gtrsim$ 10$^{-8}$ 
M$_{\odot}$/yr), such as in DG Tau and DR Tau (Hessman \& Guenther 1997). WB
detected these lines in CIDA-1 (M5.5) and GM Tau (M6.5), two Taurus objects at
the sub-stellar boundary that also exhibit broad H$\alpha$ 
profiles and veiling, and inferred accretion rates ($\sim$ 10$^{-8.5}$ 
M$_{\odot}$/yr) similar to CTTS. In fact, Muzerolle, Hartmann, \& 
Calvet (1998; hereafter MHC) found a correlation between the accretion rates 
and the line fluxes in CaII, OI and HeI for their sample of CTTS, and argued 
that the broad components of these lines originate in magnetospheric infall 
regions. MHC also found that the OI (8446\AA) line appears in emission 
only for those CTTS with accretion rates higher than a certain threshold 
value ($\sim$ 10$^{-8}$ M$_{\odot}$/yr). Thus, our detection of broad OI 
and CaII emission in IC348-165 and IC348-205 (as well as broad HeI in the 
former) is significant, and implies that some VLM objects are accreting in
a fashion analogous to CTTS. 

We also find broad OI and CaII emission in a third object, IC348-415 (M6.5) 
(Fig.2).  The results of MHC quoted above then indicate that accretion is 
very likely in this object as well.  It is thus noteworthy that its H$\alpha$ 
10\% full -width ($\sim$ 213 kms$^{-1}$) is somewhat below the accretion 
cutoff (270 kms$^{-1}$) defined by WB.  We suggest that the latter cutoff 
may need revision for VLM objects, for the following reason.  The broad 
component of the H$\alpha$ line is believed to arise in a nearly free-falling 
flow between the inner edge of the disk and the stellar surface (e.g., 
Hartmann, Hewett, \& Calvet 1994); the lower limit on the line-broadening 
is then set by the free-fall velocity ($v_{ff}$).  For ages 1--3 Myr and 
substellar masses ($M$ $\lesssim$ 0.08 M$_{\odot}$), evolutionary models 
(Baraffe et al. 1998) yield $v_{ff}$ $\sim$ 75--125 kms$^{-1}$, implying 
H$\alpha$ full-widths of order 150--250 kms$^{-1}$, significantly less than 
the WB cutoff of $\sim$ 270 kms$^{-1}$.  Muzerolle et al. (2001) argue 
that Stark broadening can increase the H$\alpha$ width beyond that expected 
from infall alone.  However, this effect does not appear important at rates 
below $\sim$ 10$^{-9}$ M$_{\odot}$/yr; thus in mildly accreting brown dwarfs, 
the H$\alpha$ line-width is likely to be set by $v_{ff}$ alone.  For instance, 
in V410 Anon 13 (a Taurus M6 object with \mdot $\sim$ 5$\times$10$^{-12}$ 
M$_{\odot}$/yr), Muzerolle et al. (2000) adequately model H$\alpha$ with 
no recourse to Stark broadening. These authors also show that accretion 
rates $\lesssim$ 10$^{-9}$ M$_{\odot}$/yr are not expected to produce any 
optical veiling arising from the accretion shock.  The WB cutoff, however, 
is {\it defined} in terms of optical veiling: only objects with veiling are 
adopted as accretors, and it is these that have H$\alpha$ full-widths $>$ 
270 kms$^{-1}$.  Our foregoing discussion suggests that VLM objects with 
very low accretion rates can fail the WB accretion test: their H$\alpha$ 
profiles are unlikely to be Stark-broadened beyond the $v_{ff}$ full-widths, 
which are less than 270 kms$^{-1}$; furthermore, they will also not exhibit 
any veiling indicative of accretion.  A case in point is V410 Anon 13 itself: 
its H$\alpha$ 10\% full-width is $\sim$ 250 kms$^{-1}$ (slightly below the WB 
limit), and it shows no veiling (and no CaII or OI emission either), yet is 
an accretor. Similarly IC348-415, which does show CaII and OI but has 
an H$\alpha$ full-width somewhat below the WB limit, is very likely to be
accreting. 

Consequently, we adopt a 10\% full-width $\sim$ 200 kms$^{-1}$ as our 
accretion cutoff.  Given the rotational velocities in our sample (all 
$v$~sin~$i$ $<$ 50kms$^{-1}$), such a width is hard to explain by 
rotational broadening and/or flaring, but is in agreement with 
the expected $v_{ff}$.  In that case, at least two other targets 
in our sample -- IC348-382 (M5.5) and IC348-355 (M8) -- are accretors 
as well, albeit with low accretion rates.  Both show broad asymmetric 
H$\alpha$ profiles (10\% full-width $>$ 200kms$^{-1}$) associated with 
infall, even though the 10\% width is somewhat below the WB limit and no 
CaII or OI emission is noticeable.  If so, IC348-355 is the latest 
spectral type (lowest mass) accretor known to date. 

Putting together now all the young objects near or below the substellar 
boundary ($\sim$ M5 and later) with published high-resolution optical 
spectra, we have 4 objects in $\rho$ Ophiuchus, 10 in IC 348, 14 in 
Taurus and 11 in Upper Scorpius\footnote{Most from this paper and 
Jayawardhana, Mohanty \& Basri (2002); 
9 in Taurus from WB and 1 from Muzerolle et al. (2000).}.  Of these, optical 
spectral signatures of accretion are found in 1 object in $\rho$ Oph (GY 5), 
5 in IC 348 (discussed above), 3 in Taurus (CIDA-1, GM Tau, V410 Anon 13) and 
1 in Upper Sco (USco 75 (M6), adopting an accretion cutoff of $\sim$ 200 
kms$^{-1}$ in 10\% width). The vast majority of $\rho$ Oph VLM objects were 
inaccessible to our optical spectroscopy because of significant extinction, 
presumably due to circumstellar as well as interstellar material.  Thus, 
our (small) $\rho$ Oph sample is heavily biased against possible accretors, 
and should not be used to estimate the accreting fraction.  Considering the 
other three clusters, which are much less affected by this bias, we find 
that $\sim$50\% of the VLM objects show disk accretion at an age $\lesssim$ 
2 Myr (IC 348), $\sim$ 20\% at age $\sim$ 3 Myr (Taurus), and $\lesssim$10\% 
by $\sim$ 5 Myr (Upper Sco).  While there are uncertainties in the cluster 
ages, IC 348 is likely to be younger than Taurus and Upper Sco. Thus, we 
appear to be seeing a decrease in the fraction of accreting young sub-stellar 
objects with increasing age.  

A decrease in brown dwarf disk frequency with age is also seen in 
measurements of infrared excess.  While a large fraction --$\sim$60--80\%-- of
brown dwarfs in the $\lesssim$ 3 Myr-old $\rho$ Oph, IC 348, and Trapezium 
clusters show near-infrared excess indicative of disks, the fraction appears 
to be lower in the somewhat older $\sigma$ Orionis ($\sim$5-8 Myr) and 
TW Hydrae ($\sim$10 Myr) groups (Jayawardhana et al. 2003; Jayawardhana, 
Ardila \& Stelzer 2002; Muench et al. 2001, Liu et al. 2003). The disk 
fractions as measured by near-infrared excess are roughly comparable 
between TTS and sub-stellar objects in the same cluster. The evidence to 
date is consistent with similar timescales for inner disk dissipation in 
young brown dwarfs and low-mass stars (Jayawardhana et al. 1999; 2001; 
Haisch, Lada \& Lada 2001). 

\subsection{Rotation and Activity}
In field $\sim$ M5--M8 dwarfs, a saturation-type rotation-activity 
connection is seen: stars rotating above a threshold 
velocity ($\sim$ 5 kms$^{-1}$) exhibit a saturated level of chromospheric 
H$\alpha$ flux, while slower rotators show a range of (smaller) fluxes 
(Mohanty \& Basri 2003).  We will explore this issue for our young VLM 
objects in a future paper, after deriving effective temperatures so that 
observed H$\alpha$ equivalent widths can be converted to fluxes.  However, 
we note here that the H$\alpha$ widths in our sample of non-accretors are 
at least as large as in saturated field dwarfs of similar spectral type 
($\sim$ 3-10\AA).  Indeed, in three cases -- IC348-256 and CFHT-BD-Tau-3 
and 4 -- the equivalent widths are many times higher than usually seen in 
quiescent M dwarfs, and are reminiscent of intense chromospheric flaring. 
Interestingly, CFHT-BD-3 shows very strong H$\alpha$ emission in two 
observations separated by a year.  All these facts indicate that 
chromospheric activity levels in young VLM objects are at least as high as, 
and possibly significantly higher than, in older field dwarfs of similar 
spectral type.  

\section{Concluding Remarks}
Our high-resolution optical spectra provide the most compelling evidence
yet for a T Tauri-like accretion phase in many sub-stellar objects, 
qualitatively similar to that of their stellar counterparts. The lifetimes
of inner disks around brown dwarfs do not appear to be vastly different from 
those of T Tauri disks. Thus, the observational evidence to date favors a 
similar mechanism for the formation of stellar and sub-stellar objects. 

While it is too early to rule out the ejected embryo hypothesis for the 
origin of at least some brown dwarfs, our results do not support it: the
naive expectation of signficantly shorter lifetimes for sub-stellar disks
is not borne out by the data. However, current numerical simulations do 
not have sufficient resolution to determine whether a sub-stellar object 
with a small, truncated disk should be an accretor at a few million years
(Bate et al. 2003). Future simulations should clarify this issue. On the 
observational front, longer-wavelength measurements with the {\it Space 
InfraRed Telescope Facility} and/or the {\it Stratospheric Observatory 
For Infrared Astronomy} could constrain the sizes of brown dwarf disks 
and thus determine if they are truncated as expected in the ejection 
scenario. Other important insights could come from investigating even 
younger objects, namely embedded (proto-) brown dwarfs, with 
high-resolution near-infrared spectroscopy.

\acknowledgements
We would like to acknowledge the great cultural significance of Mauna Kea 
for native Hawaiians, and express our gratitude for permission to observe 
from its summit. We also thank the Keck Observatory staff 
for their outstanding assistance over the past several years. This work was 
supported in part by NSF grants AST-0205130 to R.J. and AST-0098468 to G.B. 
S.M. acknowledges the support of a SIM-YSO postdoctoral fellowship.


\clearpage
\begin{figure}
\epsscale{0.75}
\plotone{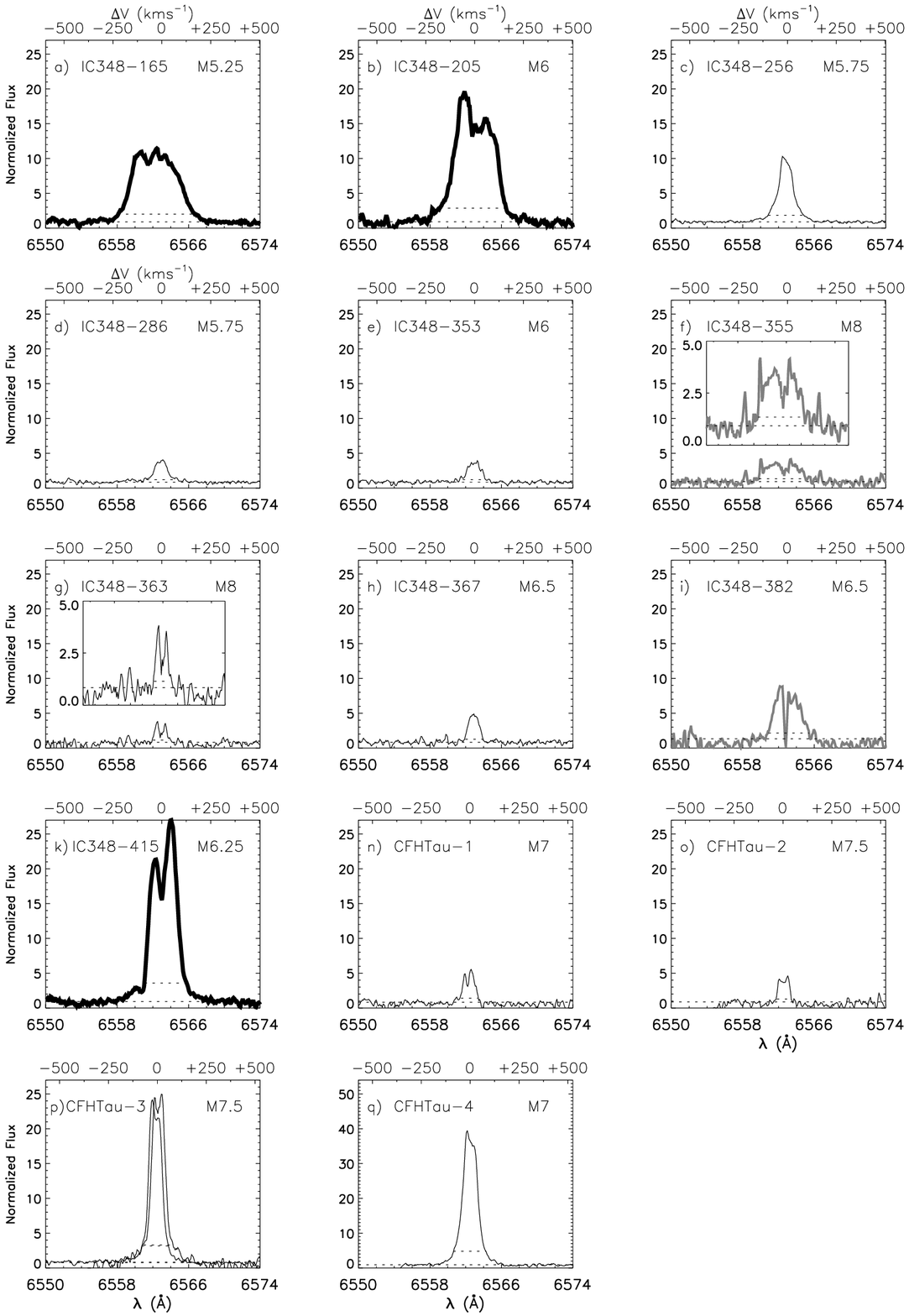}
\caption{H$\alpha$ line profiles of the target sample. Spectra shown have
been smoothed by a 3-pixel boxcar; continuum and full width at 10\% of the 
peak levels are marked by dotted lines.  Thick black lines indicate accretors 
with broad H$\alpha$ as well as CaII and OI emission; grey indicates probable 
accretors, based on the H$\alpha$ profile-shape and 10\% full-width.  Insets 
zoom in on objects with low peak-flux and noisy continuua, to clearly show 
the H$\alpha$ detection.  For CFHT-3, two spectra are shown, separated by a 
year; note the similarly strong emission both times.  For CFHT-4, note the 
change in Y-axis scale; the peak flux in this object is {\it much} higher 
than in any other target in our sample.}
\end{figure}

\clearpage
\begin{figure}
\epsscale{0.5}
\plotone{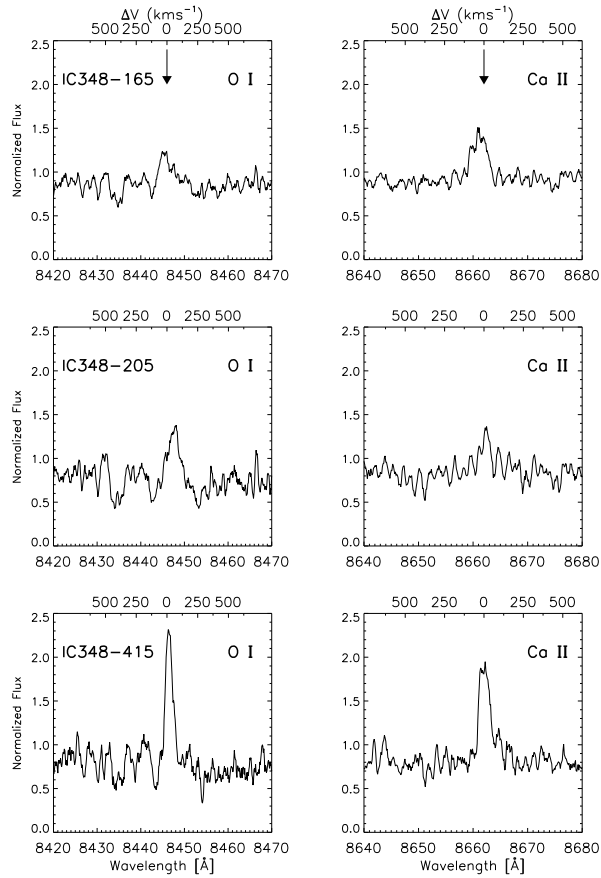}
\caption{Line profiles of OI (8446\AA) and CaII (8662\AA) in three IC 348 
objects.  Arrows indicate line-center.  These spectra have been smoothed 
by a 4-pixel boxcar.}
\end{figure}

\clearpage
\bigskip
\begin{center}
\begin{deluxetable}{lcccccc}
\tablecaption{\label{tab1}}
\tablehead{
\colhead{Object} &
\colhead{Sp Type\tablenotemark{a}} &
\colhead{v~{\it sin}~i} &
\colhead{H$\alpha$ EW\tablenotemark{b}} &
\colhead{H$\alpha$ 10\% FW} &
\colhead{OI EW\tablenotemark{b, c}} &
\colhead{CaII EW\tablenotemark{b, c}} \\
 & & (kms$^{-1}$) & (\AA) & (kms$^{-1}$) & (\AA) & (\AA) \\}

\startdata
IC348-165  &M5.25   &19$\pm$2        &66$\pm$6       & 389  & 1.1  & 1.8 \\
IC348-205  &M6   &6$\pm$2         &93$\pm$9       & 338  & 2.4  & 1.5: \\
IC348-256  &M5.75   & 9$\pm$2        &23$\pm$2       & 180  & -  &  - \\
IC348-286  &M5.75   &19 $\pm$3       & 6.9$\pm$0.7       & 148  & -  &  -  \\
IC348-353  &M6   &25$\pm$3        & 5.8 $\pm$0.6      & 117   & -  &  - \\
IC348-355  &M8   &45:             & 10.3$\pm$1      & 235   & -  &  - \\
IC348-363  &M8   &14$\pm$3        & 3.8:      &  82:  & -  &  - \\
IC348-367  &M5.75 &20$\pm$3        & 6.8:      &  92:  & -  &  - \\
IC348-382  &M5.5 &10:       &15$\pm$2       & 208   & -  &  - \\
IC348-415  &M6.5   &$<$5        &80$\pm$8       & 213  & 3.6  & 3.0 \\
CFHT-BD-Tau-1   &M7   &7$\pm$3         & 7.4$\pm$1.5       & 85   & -  &  - \\
CFHT-BD-Tau-2   &M7.5   &8:        & 11$\pm$1       & 80   & -  &  - \\
CFHT-BD-Tau-3   &M7.75  &12$\pm$2  & 50$\pm$4/65$\pm$1 & 136/153   & -  &  - \\
CFHT-BD-Tau-4   &M7   &11:       &  69$\pm$4       & 175  & -  &  - \\
\enddata
\tablenotetext{a}{Spectral types for IC 348 sources are from Luhman et al. 
(2003; for 367, 382 and 415) and Luhman (1999; for others). Spectral type for 
CFHT-BD-Tau-1 is from Mart\'\i n et al. (2001).  Spectral types for 
remaining CFHT Taurus sources are from Brice\~{n}o et al. (2002), 
who use the same classification scheme as Luhman (1999) and somewhat 
different from Mart\'\i n et al. (2001). For CFHT-2 and 3, their spectral 
types are respectively 0.5 and 1.25 subclasses earlier than those of 
Mart\'\i n et al. (2001).} 
\tablenotetext{b}{{\it Pseudo}-equivalent width; the plethora of molecular 
lines make it impossible to determine the true continuum in these stars.}
\tablenotetext{c}{For OI and CaII, large absorption troughs near the emission 
lines (see Fig. 2) make line-width determination difficult; our errors are 
$\sim$ $\pm$15\%.}
\end{deluxetable}
\end{center}


\begin{references}
\reference{} Baraffe, I., et al. 1998, A\&A, 337, 403
\reference{} Basri, G., et al. 2000, ApJ, 538, 363
\reference{} Bate, M.R., Bonnell, I.A., \& Bromm, V. 2002, MNRAS, 332, L65
\reference{} Bate, M.R., Bonnell, I.A., \& Bromm, V. 2003, MNRAS, 339, 577
\reference{} Brice\~{n}o, C., et al. 2002, ApJ, 580, 317
\reference{} Haisch, K.E.,Jr., Lada, E.A., \& Lada, C.J. 2001, 553, L153
\reference{} Hartmann, L., Hewett, R., \& Calvet, N. 1994, ApJ, 426, 669
\reference{} Hessman, F.V. \& Guenther, E.W. 1997, A\&A, 321, 497
\reference{} Jayawardhana, R. et al. 1999, ApJ, 521, L129
\reference{} Jayawardhana, R. et al. 2001, ApJ, 550, L197
\reference{} Jayawardhana, R., Ardila, D.R., \& Stelzer, B. 2002, in {\it 
Brown Dwarfs}, ed. E. L. Mart\'in, San Francisco: Astronomical Society of the 
Pacific
\reference{} Jayawardhana, R., Mohanty, S., \& Basri, G. 2002, ApJ, 578, L141 
\reference{} Jayawardhana, R. et al. 2003, AJ, submitted 
\reference{} Joy, A.H. 1945, ApJ, 102, 168 
\reference{} Liu, M.C., et al. 2003, ApJ, 585, 372
\reference{} Luhman, K.L. 1999, ApJ, 525, 466
\reference{} Luhman, K.L., et al. 2000, ApJ, 540, 1016
\reference{} Luhman, K.L., et al. 2003, ApJ, submitted
\reference{} Mart\'\i n, E.L., et al. 2001, ApJ, 561, L195
\reference{} Mohanty, S. \& Basri, G., 2003, ApJ, 583, 451
\reference{} Mohanty, S., et al., 2003a, ApJ, submitted
\reference{} Mohanty, S., Jayawardhana, R., Basri, G., 2003b, ApJ, submitted
\reference{} Muench, A.A., et al. 2001, ApJ, 558, L51
\reference{} Muzerolle, J., Hartmann, L., \& Calvet, N. 1998, AJ, 116, 455 
[MHC]
\reference{} Muzerolle, J., et al. 2000, ApJ, 545, L141
\reference{} Muzerolle, J., Calvet, N. \& Hartmann, L. 2001, ApJ, 550, 944
\reference{} Padoan, P. \& Nordlund, A. 2003, ApJ, in press 
\reference{} Reipurth, B. \& Clarke, C. 2001, AJ, 122, 432
\reference{} Vogt, S.S., et al. 1994, Proc. SPIE, 2198, 362
\reference{} White, R. J., \& Basri, G. 2003, ApJ, 582, 1109 [WB]
\reference{} White, R.J. \& Ghez, A.M. 2001, ApJ, 556, 265 
\end{references}
\end{document}